\RequirePackage{cmap}%
\documentclass{scrartcl}
\pdfoutput=1

\usepackage[utf8]{inputenc}
\usepackage[T1]{fontenc}
\usepackage{ifpdf}
\usepackage{cite}
\usepackage{caption}

\usepackage[charter]{mathdesign}
\usepackage{berasans,beramono}
\usepackage[final,tracking=true,kerning=true,spacing=true]{microtype}
\microtypecontext{spacing=nonfrench}

\usepackage{textcomp}
\usepackage{amsmath}

\usepackage{booktabs}

\newcommand*{\mleft}{\mathopen{}\mathclose\bgroup\left}
\newcommand*{\mright}{\aftergroup\egroup\right}
\newcommand*{\chTime}[1]{\Delta\tau_{\mathrm{ch}}\mleft(#1\mright)}
\newcommand*{\fpTime}[2]{\Delta\tau_{\mathrm{fp}}\mleft(#1,#2\mright)}
\newcommand*{\nthStopOn}[2]{#1\textrm{\scriptsize @}#2}
\newcommand*{\arrTime}[2]{\tau_{\mathrm{arr}}\mleft(\nthStopOn{#1}{#2}\mright)}
\newcommand*{\depTime}[2]{\tau_{\mathrm{dep}}\mleft(\nthStopOn{#1}{#2}\mright)}
\newcommand*{\seq}[1]{\mleft<#1\mright>}
\newcommand*{\stopsOn}[1]{\vec{s}\mleft(#1\mright)}
\newcommand*{\lineOf}[1]{L\mleft(#1\mright)}
\newcommand*{\transfer}[4]{\nthStopOn{#1}{#2} \rightarrow \nthStopOn{#3}{#4}}
\newcommand*{\cutNode}{N_\mathrm{cut}}
\newcommand{\corresponds}{\overset{\scriptscriptstyle\wedge}{=}}

\ifpdf
\fi

\begin{document}
\title{Trip-Based Public Transit Routing Using Condensed Search Trees}
\author{Sascha Witt \\ \texttt{sascha.witt@kit.edu} \\ \\
  Karlsruhe Institute of Technology (KIT)\\Karlsruhe, Germany}
\date{}
\maketitle

\begin{abstract}
We study the problem of planning Pareto-optimal journeys in public transit networks.
Most existing algorithms and speed-up techniques work by computing subjourneys to intermediary stops until the destination is reached.
In contrast, the trip-based model~\cite{Witt2015} focuses on trips and transfers between them, constructing journeys as a sequence of trips.
In this paper, we develop a speed-up technique for this model inspired by principles behind existing state-of-the-art speed-up techniques, Transfer Patterns~\cite{Bast2010} and Hub Labelling~\cite{Cohen2003}.
The resulting algorithm allows us to compute Pareto-optimal (with respect to arrival time and number of transfers) 24-hour profiles on very large real-world networks in less than half a millisecond.
Compared to the current state of the art for bicriteria queries on public transit networks, this is up to two orders of magnitude faster, while increasing preprocessing overhead by at most one order of magnitude.
\end{abstract}

\section{Introduction}\label{sec:introduction}
Finding optimal journeys in public transit networks is a complex problem.
Efficient algorithms are required to allow real-time answering of queries by users in online systems such as Google Maps Transit\footnote{\texttt{https://maps.google.com/transit}} or those of local providers such as \texttt{bahn.de} or \texttt{fahrplan.sbb.ch}.
In these systems, users enter a source location, a destination, and a rough point in time and expect a number of journeys that are optimal in some sense.

Precisely what constitutes an optimal journey is non-trivial to define, as it often depends on individual user preferences.
Generally, passengers want to arrive as quickly as possible, so travel time should usually be minimized.
However, some users may prefer a slightly longer journey with fewer transfers between different vehicles, as transfers reduce travel comfort and introduce additional uncertainty --- connecting trains might be missed due to delays.
How much extra travel time someone is willing to accept in exchange for fewer transfers differs from user to user and might depend on several factors, such as arrival time or even purpose of the journey.

Since no system can capture all these variables to compute the optimal journey for each query, we usually compute a set of possible journeys and let the user choose among them, possibly after applying some filtering~\cite{Delling2013,Farina2004}.
A general approach to this is to define a number of criteria, such as arrival time and number of transfers, and compute a set of Pareto-optimal journeys, i.e.\ a set such that no journey is better than any other in all criteria.

\subsection{Related Work}\label{sec:related_work}
In the past, several algorithms based on different principles have been proposed.
For an extensive survey, please refer to Bast et~al.~\cite{Bast2014a}.
Pyrga et~al.~\cite{Pyrga2008} reduce the problem of finding optimal journeys in public transit networks to finding shortest paths in graphs.
They propose the time-extended and time-dependent model along with some speed-up techniques such as goal directed search, and optimize both travel time and number of transfers in the Pareto sense.
Geisberger~\cite{Geisberger2010} applies the concept of contraction hierarchies, which have proved successful on road networks, to public transit networks.
Only travel time is optimized.
Berger et~al.~\cite{Berger2010} introduce SUBITO and k-flags, two speed-up techniques that optimize both travel time and number of transfers in the Pareto sense.

RAPTOR~\cite{Delling2012} foregoes modelling the data as a graph and instead operates directly on the timetable data.
In addition to travel time and number of transfers, they also consider price as a criteria.
The Connection Scan Algorithm (CSA)~\cite{Dibbelt2013} also eschews graphs and instead works on an ordered array of connections to find Pareto-optimal journeys with respect to travel time and number of transfers.
Accelerated CSA~\cite{Strasser2014} is a speed-up technique for CSA that works via partitioning of the network.
Unlike the original CSA, it was only evaluated as a single-criterion algorithm, using the number of transfers as a tiebreaker between journeys with identical arrival time.

Public Transit Labelling (PTL)~\cite{Delling2015} uses, as the name implies, a hub labelling approach.
It requires extensive preprocessing and produces a very large amount of auxiliary data, but leads to very low query times, even for multi-criteria queries.
Timetable Labelling (TTL)~\cite{Wang2015} is another labelling-based approach, which has been extended in the context of databases by Efentakis~\cite{Efentakis2016}.
However, TTL only performs single-criterion queries regarding arrival time.

Transfer Patterns (TP)~\cite{Bast2010,Bast2016,Bast2014} is a speed-up technique that precomputes the eponymous transfer patterns between all stops in the network.
These transfer patterns are formed by the sequence of stops where passengers transfer between vehicles.
At query time, these patterns are then used to quickly find all Pareto-optimal journeys.

\subsection{Our Contribution}\label{sec:contribution}
In this work, we present a speed-up technique based on Trip-Based Public Transit Routing (TB)~\cite{Witt2015}.
Unlike other approaches, TB conceptually works on a graph where nodes represent trips, not stops.
Edges represent possible transfers between trips, and are qualified using the indices of the stops where passengers exit or board a trip.
These transfers are precomputed and can be looked up quickly during query processing.
This has the advantage that minimum change times and footpaths do not have to be evaluated at query time, and allows fine-grained modelling without query-time overhead.

Inspired by the principles behind Transfer Patterns~\cite{Bast2010} and Hub Labelling~\cite{Cohen2003}, our speed-up technique achieves sub-millisecond query times for profile queries on country-sized networks, while keeping preprocessing overhead low.

\section{Preliminaries}\label{sec:preliminaries}
A public transit network is defined by an aperiodic \emph{timetable}, which contains a set of stops, a set of footpaths, and a set of trips.
A \emph{stop} is a physical location where passengers can enter or exit a vehicle, such as a bus or train.
Depending on the granularity of the model, a stop may represent an entire train station, a single platform, or some subset of all platforms within a train station.
Transferring from one vehicle to another at the same stop $s$ may require a certain amount of time, which we call minimum change time $\chTime{s}$.
If the time between the arrival of the previous vehicle and the departure of the subsequent one is less than $\chTime{s}$, no transfer between them is possible at this station.
\emph{Footpaths} connect two stops and indicate the time required to walk from one to the other.
We use the most general model of directed, non-transitive footpaths.
We denote the time required to walk from stop $s_1$ to $s_2$ as $\fpTime{s_1}{s_2}$, with $\fpTime{s_1}{s_2} := \infty$ if no footpath from $s_1$ to $s_2$ exists.

\emph{Trips} represent vehicles.
Each trip $t$ travels along a sequence of stops $\stopsOn{t} = \seq{\nthStopOn{t}{1},\dots,\nthStopOn{t}{n}}$.
A trip may visit a stop multiple times.
For each $\nthStopOn{t}{i}$, the timetable contains the arrival time time $\arrTime{t}{i}$ and the departure time $\depTime{t}{i}$ of trip $t$ at that stop index.
Trips that travel along the same sequence of stops are grouped into \emph{lines}.
We require that trips never overtake another trip of the same line; more specifically, we require that the trips of a line can be totally ordered with respect to
\begin{equation}
  t_1 \preceq t_2 \iff \forall i \in \mleft[1, \mleft|\stopsOn{t_1}\mright|\mright]:
    \arrTime{t_1}{i} \leq \arrTime{t_2}{i}\text{.}
\end{equation}
Trips that violate this requirement are assigned to different lines.
We denote the line of a trip $t$ as $\lineOf{t}$, and define $\stopsOn{\lineOf{t}} := \stopsOn{t}$.

\emph{Transfers} indicate connections between trips.
We denote transfers as $\transfer{t_1}{e}{t_2}{b}$, meaning that passengers can exit trip $t_1$ at stop index $e$ in order to board trip $t_2$ at stop index $b$.
Transfers may occur at a single stop, in which case
\begin{equation}
  \arrTime{t_1}{e} + \chTime{\nthStopOn{t_1}{e}} \leq \depTime{t_2}{b}
\end{equation}
must hold, or they may involve a footpath, in which case the requirement is
\begin{equation}
  \arrTime{t_1}{e} + \fpTime{\nthStopOn{t_1}{e}}{\nthStopOn{t_2}{b}}
    \leq \depTime{t_2}{b}\text{.}
\end{equation}
A \emph{journey} describes how and when to get from a source stop $s_\mathrm{src}$ to a destination stop $s_\mathrm{dest}$.
It can be defined by a sequence $\seq{\nthStopOn{t_1}{b_1},\nthStopOn{t_2}{b_2},\dots,\nthStopOn{t_n}{b_n}}$ with the following requirements:
\begin{gather}
  s_\mathrm{src} = \nthStopOn{t_1}{b_1}
    \lor \fpTime{s_\mathrm{src}}{\nthStopOn{t_1}{b_1}}
    < \infty\label{eq:journey_fp_src}\\
  \forall i \in \mleft[1,n\mright): \exists e > b_i:
    \transfer{t_i}{e}{t_{i+1}}{b_{i+1}}\label{eq:journey_transfers}\\
  \exists i: \nthStopOn{t_n}{i} = s_\mathrm{dest} \lor
    \fpTime{\nthStopOn{t_n}{i}}{s_\mathrm{dest}} < \infty\label{eq:journey_fp_dest}
  \text{.}
\end{gather}
These requirements ensure that the first trip can be reached \eqref{eq:journey_fp_src}, that transfers are possible between subsequent trips \eqref{eq:journey_transfers}, and that the final trip arrives at the destination \eqref{eq:journey_fp_dest}.

We consider two well-known bicriteria problems, optimizing arrival time and number of transfers required.
It has been shown~\cite{Muller-Hannemann2006} that for these criteria, computing the full set of Pareto-optimal journeys is feasible.
A journey is Pareto-optimal if no other journey dominating it exists.
A journey dominates another if it is better or equal in all criteria --- if they are equal in all criteria, we break the tie arbitrarily and keep only one of them in the set.

The input to the \emph{earliest arrival query} consists of a source stop $s_\mathrm{src}$, a destination stop $s_\mathrm{dest}$, and a departure time $\tau$.
The result is a set of tuples $\mleft(\tau_\mathrm{dest}, n\mright)$, one for each Pareto-optimal journey leaving $s_\mathrm{src}$ no earlier than $\tau$ and arriving at $s_\mathrm{dest}$ at time $\tau_\mathrm{dest}$ after $n$ transfers.
For the \emph{profile query}, we are given a source stop $s_\mathrm{src}$, a destination stop $s_\mathrm{dest}$, an earliest departure time $\tau_\mathrm{edt}$, and a latest departure time $\tau_\mathrm{ldt}$.
Here, we consider the departure time of journeys as an additional criterion, with later departures dominating earlier ones.
Thus, we compute all Pareto-optimal journeys departing at $s_\mathrm{src}$ at some time $\tau_\mathrm{src}$ with $\tau_\mathrm{edt} \leq \tau_\mathrm{src} \leq \tau_\mathrm{ldt}$ and arriving, after $n$ transfers, at $s_\mathrm{dest}$ at time $\tau_\mathrm{dest}$.
The answer to the query is then the set of tuples $\mleft(\tau_\mathrm{src},\tau_\mathrm{dest},n\mright)$ corresponding to these journeys.

\section{Trip-Based Public Transit Routing}\label{sec:tbptr}
This section provides a quick explanation of the Trip-Based Public Routing (TB) algorithm~\cite{Witt2015}.
For more details, please refer to the original publication.

\subsection{Preprocessing}
As mentioned in section~\ref{sec:contribution}, TB uses trips and transfers between them as its basic building blocks.
During a short preprocessing phase, all possible transfers between trips are computed.
However, it can be shown that many of these transfers can never be part of an optimal journey, for example transfers that lead to trips that run in the opposite direction, or transfers to several trips of the same line.
Therefore, the second step of preprocessing is discarding these superfluous transfers, which may constitute up to 90\% of total transfers.
Since each trip can be processed separately, preprocessing is trivially parallelized and can be performed within minutes even for very large networks.

\subsection{Queries}\label{sec:tb_queries}
Queries are similar to a breadth-first search on the graph formed by trips and the transfers between them.
For an earliest arrival query, we first identify the trips reachable from the source stop, and insert them into a queue.
Then, each trip is processed by scanning its outgoing transfers.
Newly reached trips are in turn added to the queue.
Trips are marked as reached by, conceptually, assigning labels consisting of trip, stop index, and number of transfers needed to reach that trip to lines.\footnote{In the implementation, we unroll the ``trip'' and ``number of transfers'' dimensions for faster lookup and to allow the use of SIMD instructions.}
Branches of the search are pruned if they are dominated by existing labels.
The graph is explored until all Pareto-optimal journeys to the destination stop are found.

For a profile query, we essentially repeat the above multiple times.
Observe that the departure time is an additional criteria for journeys in profile queries, with later journeys dominating earlier ones.
If all other criteria are equal, the journey with the later departure has less travel time and is therefore preferable.
Thus, we start with the latest possible departure at the source stop, and perform an earliest arrival query.
We can add the resulting journeys to the result set.
Then, without resetting labels, we perform an earliest arrival query for the second-latest departure, and so on.
By preserving the labels between runs, we allow later journeys to dominate earlier ones, avoiding redundant work.

\section{Storing One-to-All Search Trees}\label{sec:algorithm}
In this section, we show how some of the principles behind Transfer Patterns~\cite{Bast2010} can be applied to the Trip-Based model.
The core idea of the Transfers Patterns algorithm is to precompute, for all pairs of source and destination stop, the \emph{transfer patterns} for all optimal journeys.
The transfer pattern of a journey is the sequence of stops where a change of vehicle occurs.
In practice, optimal journeys between two given stops share a limited number of transfer patterns.
If all optimal transfer patterns between source and destination are known, queries can be answered quickly by looking up direct connections between the stops forming the transfer patterns.

We use the same property as the foundation for our speed-up technique.
Since we operate on trips --- or more generally, lines, which are ordered sets of trips --- we do not precompute sequences of stops where transfers occur.
Instead, we precompute the sequence of \emph{lines} that correspond to an optimal journey, together with the stop indices where each of these lines is boarded.
As we show in the next section, these line sequences form a natural generalization of one-to-all profile search trees.

\subsection{Prefix Trees}\label{sec:prefix_trees}
We compute one-to-all profiles from each stop to find all potential line sequences.
These one-to-all profiles are at their core identical to the one-to-one profiles described in the original publication~\cite{Witt2015} and summarized in section~\ref{sec:tb_queries}.

First, all departures at the source stop are ordered by departure time and then processed backwards.
For each distinct departure time, we then perform a breadth-first search as described in section~\ref{sec:tb_queries}.
This results in a breadth-first tree, with the source stop as the root node, the visited trips as internal nodes, and the reached stops as leaves.
In contrast to one-to-one profiles, we also assign labels to all stops, consisting of arrival time and number of transfers.
We update these using the breadth-first tree, pruning branches that do not lead to an improved stop label.
The remaining tree is generalized by replacing all trips with their respective line and the index of the stop where the trip was boarded.
We then restart the search using the next (earlier) departure, preserving all labels.

Finally, the trees are merged, resulting in one \emph{prefix tree}~\cite{DeLaBriandais1959} for each source stop, containing the optimal line sequences to all destination stops.
In essence, this prefix tree represents a condensed, time-independent result of a one-to-all profile search.
Note that prefix trees are functionally equivalent to the transfer pattern graphs used by Transfer Patterns~\cite{Bast2010}, except that internal nodes represent lines instead of stops.

\subsection{Queries}\label{sec:queries}
We can use these prefix trees to quickly answer queries.
First, we construct the \emph{query graph}.
To do so, we find the nodes corresponding to the destination stop in the prefix tree of the source stop.
We follow the paths from these nodes to the root, adding edges from parent to child nodes to the query graph.
Multiple occurrences of the same $\nthStopOn{L}{b}$ in the prefix tree are mapped onto the same node in the query graph.
Again, note the similarity to the query graph used by Transfer Patterns~\cite{Bast2010}.

To answer the query, we run a simple multi-criteria label-correction shortest path algorithm~\cite{Disser2008} on the query graph.
Labels consist of a trip, the number of transfers, and, for profile queries, the departure time at the source stop.
Finding the initial trips at the source stop is straightforward.
Given a label $(t, n, \tau_{dep})$, we relax an edge between $\nthStopOn{L_1}{i}$ and $\nthStopOn{L_2}{j}$ by finding a transfer $\transfer{t}{k}{s}{j}$ such that $k > i$ and $\lineOf{s} = L_2$.
We then add a label $(s, n+1, \tau_{dep})$ to the node representing $\nthStopOn{L_2}{j}$.
Once the algorithm terminates, we can extract the arrival times at the destination stop from the labels.
Intuitively, the prefix tree tells us which paths through the networks optimal journeys can take.
The query then follows these paths to find the actual journeys for the given departure time(s).

\section{Splitting Trees}\label{sec:splitting_trees}
Unfortunately, for large networks, prefix trees grow unfeasibly large, and memory usage becomes an issue.
Each tree spans the entire network, and in addition, many subtrees are duplicates of each other, with slightly different prefixes.
Furthermore, subtrees are often duplicated across different trees, since stops can only be reached through a limited set of lines.

We can reduce this redundancy by removing branches from the prefix trees and instead storing them in \emph{postfix trees}.
These postfix trees are essentially reverse prefix trees: They are rooted at a \emph{destination} stop and describe optimal line sequences for reaching that stop.
Storing these sequences once for each destination stop instead of once or even multiple times for each source stop greatly improves space efficiency.
Optimal line sequences can be recovered by concatenating branches of the source's prefix tree with matching branches of the destination's postfix tree.

\subsection{Postfix Trees}\label{sec:postfix_trees}
We construct the postfix trees from the prefix trees as follows.
For each path from the root (that is, the source stop) to a leaf (a destination stop), we select an internal node $\cutNode$ where we ``cut'' this path.
Section~\ref{sec:cut_selection} explains how this node is chosen.
We add the subpath from $\cutNode$ (inclusive) to the leaf --- in reverse order --- to the postfix tree for the destination node.
Then, we remove the leaf node and, recursively, all internal nodes that no longer have any children from the prefix tree, until we reach $\cutNode$.
Thus, if the prefix tree originally contained the path $\seq{S,N_1,\dots,N_l,\cutNode,N_{l+1},\dots,N_n,T}$, we end up with $\seq{S,N_1\dots,N_l,\cutNode}$ in the prefix tree and $\seq{T,N_n,\dots,N_{l+1},\cutNode}$ in the postfix tree.

However, recall that each internal node represents a line together with the stop index where the line is boarded, $\nthStopOn{L}{b}$.
This breaks the symmetry between prefix and postfix trees.
As a result, we end up with many postfixes that are identical except for the board index at $\cutNode$ --- depending on the source stop, there are many ways to reach a line, but only a limited number of (optimal) ways from that line to the destination stop.
We can merge these nodes by setting the index to the \emph{exit} of the next transfer, which is identical for all of them.
Note that we only do this for the $\cutNode$ in postfix trees, not for any other nodes in either prefix or postfix trees.
Thus, if $\cutNode \corresponds \nthStopOn{L}{b}$ and the original path was $\seq{S,N_1,\dots,\nthStopOn{L}{b},\dots,N_n,T}$, we now have $\seq{S,N_1,\dots,\nthStopOn{L}{b}}$ in the prefix tree for $S$ and $\seq{T,N_n,\dots,\nthStopOn{L}{e}}$ in the postfix tree for $T$, with $b < e$.
This results in a greatly reduced number of leaves in postfix trees, while still allowing us to recover the original line sequence.

Since we no longer store destination stops in prefix trees (or source stops in postfix trees), but still want to preserve directional information, a bit vector is stored with each $\cutNode$.
We partition the stops and set the $i$th bit if $\cutNode$ connects to the postfix tree of a stop in partition $i$, and vice versa for the $\cutNode$ in postfix trees.
In practice, we use 64-bit integers and simply partition the stops by ID, taking advantage of the pre-existing locality in the data sets.

\subsection{Queries}\label{sec:split_query_graph}
The algorithm for query graph construction follows from the construction of the postfix trees.
First, we take the prefix tree for the source stop and select all $\cutNode$ where the bit vector has the bit corresponding to the destination stop set.
Similarly, we select the $\cutNode'$ from the postfix tree for the destination stop where the bit corresponding to the source stop is set.
Then, we find all pairs $(\cutNode,\cutNode')$ such that $\cutNode \corresponds \nthStopOn{L}{b}$ and $\cutNode' \corresponds \nthStopOn{L}{e}$ with $b < e$.
Each such pair defines a path $\seq{S,N_1,\dots,N_l,\cutNode,N_{l+1},\dots,N_n,T}$, and we need to ensure that the query graph contains all edges in that path.
By ordering the nodes by their corresponding line, we can find these pairs using an algorithm similar to a coordinated sweep.
Due to the generalizations performed during postfix tree construction, we will find some prefix-postfix combinations that do not correspond to an optimal line sequence.
Thus, the resulting query graph will usually be larger than in section~\ref{sec:queries}, but this only affects performance of the query, not correctness.
The query algorithm itself is the same as before.

Essentially, we find optimal paths during preprocessing, split them at some intermediary node for more efficient storage, and then reassemble them at query time.
Note the similarity to the concept of hub labelling~\cite{Cohen2003}.
In hub labelling, optimal journeys are split at some intermediary hub, then stored in compressed form at the source and destination.
We do the same, except we only store the more general line sequences instead of the journeys, which we can then reconstruct at query time.
Indeed, as we show in section~\ref{sec:experiments}, our approach shares some properties with existing labelling approaches.

The flags we use to filter possible connections are reminiscent of arc flags~\cite{Mohring2007}.
Without them, many long prefixes would connect to long postfixes for stops that are close together on the network, without a corresponding optimal journey.
Exploring these unnecessary nodes during the query would be costly and is avoided by this pre-filtering.

\subsection{Cut Selection}\label{sec:cut_selection}
It is clear that the choice of $\cutNode$ has a large effect on the resulting trees.
In general, we want smaller trees, which are more space efficient.
We examined two fundamentally different strategies.

The first is to simply cut paths in half.
Unsurprisingly, this results in rather large trees, since paths are cut at more or less arbitrary lines.
This results in many different prefixes and postfixes for each stop, which translates to large trees.

The second strategy exploits the underlying network's structure by selecting the most ``important'' lines.
To find these lines, we construct the \emph{line graph}~\cite{Bermond1977} of the network.
In the (undirected) line graph, nodes correspond to lines, and two nodes share an edge if and only if a transfer between these lines is possible.
We then use this line graph to compute the betweenness centrality~\cite{Freeman1977} of each line using Brandes' algorithm~\cite{Brandes2001}.\footnote{We chose this algorithm for simplicity; since the exact centrality is not required, one could also use an approximate algorithm~\cite{Brandes2007} instead.}
This gives us an ordering of the lines, and when choosing $\cutNode$, we select the node which corresponds to the most central line on the path.
This ensures that the choice is consistent across different paths, which allows better merging of prefixes and postfixes.
As we show in section~\ref{sec:experiments}, this strategy gives good results on country-sized networks, which typically exhibit good structure.
Unfortunately, it is less successful on the less structured metropolitan networks.
On these, using the simpler strategy of cutting paths into two equal halves leads to better results.
Exploration of further criteria for selecting cut nodes is a subject of future research.

\section{Experiments}\label{sec:experiments}
We performed experiments using a quad $8$-core Intel Xeon E5-4640 clocked at $2.4$\,GHz with $512$\,GB of DDR3-1600 RAM, using $64$ threads for parallel preprocessing.
Except where otherwise noted, computations are sequential.
Code was written in C++ and compiled using g++ 5.2.0 with optimizations enabled.
We consider five real-world data sets, three covering countries of varying size and two metropolitan networks:
Germany, provided to us by Deutsche Bahn, Switzerland, available at \texttt{gtfs.geops.ch}, and Sweden, available at \texttt{trafiklab.se}, contain both long-distance and local transit, and cover two consecutive days to allow for overnight journeys.
London, available at \texttt{data.london.gov.uk}, and Madrid, available at \texttt{emtmadrid.es}, cover a single day only.
For Madrid, we computed footpaths using a known heuristic~\cite{Delling2010}, for all other instances, they are part of the input.
These data sets are summarized in Table~\ref{tab:data_sets}.

\begin{table}[tp]
  \centering
  \caption{Instances used for experiments.}\label{tab:data_sets}
  \begin{tabular}{l r r r r r r}
    \hline
    Instance & Stops & Conn.& Trips & Lines & Footp.& Transfers \\\hline
    Germany & $247.9$\,k & $27\,061$\,k & $1\,432$\,k & $192.8$\,k & $98.8$\,k & $84\,950$\,k \\
    Sweden & $50.7$\,k & $6\,054$\,k & $261$\,k & $17.6$\,k & $0.8$\,k & $16\,455$\,k \\
    Switzerland & $27.8$\,k & $4\,650$\,k & $611$\,k & $14.4$\,k & $34.3$\,k & $12\,626$\,k \\
    London & $20.8$\,k & $4\,991$\,k & $129$\,k & $2.2$\,k & $27.6$\,k & $15\,883$\,k \\
    Madrid & $4.6$\,k & $5\,280$\,k & $190$\,k & $1.4$\,k & $1.4$\,k & $9\,256$\,k \\
    \hline
  \end{tabular}
\end{table}

Preprocessing figures can be found in Table~\ref{tab:preprocessing}.
Due to scheduling conflicts, sequential preprocessing of the Germany instance was performed on a different machine%
\footnote{Dual $8$-core Intel Xeon E5-2650s\,v2, $2.6$\,GHz, $128$\,GB DDR3-1600 RAM,  $20$\,MB L3 cache}.
We report the total time required to perform the computation of prefix and postfix trees, as described in section~\ref{sec:splitting_trees}.
This includes the time required to compute the betweenness centrality, which is negligible in most cases.
For Germany, Switzerland and Sweden we use the betweenness centrality to select cut nodes;
for London and Madrid we use the simpler method of cutting paths in half.
The reverse generally leads to larger trees and therefore higher memory consumption.
For most instances, the difference is about $1$--$2$\,GB; for Germany, the difference is almost $50$\,GB.
It is interesting to note that the metropolitan networks require more space than the two small country-sized networks.
This indicates that the topology of the network is more important than the raw size in terms of stops or connections.
A similar effect can be seen in the labelling approaches, Public Transit Labelling (PTL)~\cite{Delling2015} and Timetable Labelling (TTL)~\cite{Wang2015}.

\begin{table}[tp]
  \centering
  \caption{Preprocessing figures.
    Listed are the average time required to compute the full prefix tree for a stop,
    the total time required to compute the split trees for all stops (sequential and parallel),
    the average number of nodes in those trees (per stop, i.e.\ the sum of prefix and postfix),
    and the total space consumption.
    Sequential preprocessing for the Germany instance was performed on a different machine.
  }\label{tab:preprocessing}
  \begin{tabular}{l r r r r r r}
    \hline
    \,       & \multicolumn{1}{c}{p.\ prefix} & \multicolumn{1}{c}{seq.} & \multicolumn{1}{c}{par.} & \multicolumn{1}{c}{speed} & \multicolumn{1}{c}{avg.\ \#} & \multicolumn{1}{c}{mem.} \\
    Instance & \multicolumn{1}{c}{tree [ms]} & \multicolumn{1}{c}{[h:m]} & \multicolumn{1}{c}{[h:m]} & \multicolumn{1}{c}{up} & \multicolumn{1}{c}{of nodes} & \multicolumn{1}{c}{[GB]} \\\hline
    Germany & $2\,143.6$ & $(231$:$16)$ & $13$:$48$ & $(16.8$\,x$)$ & $7\,305$ & $23.2$ \\
    Sweden & $166.7$ & $4$:$33$ & $0$:$18$ & $15.2$\,x & $2\,433$ & $1.6$ \\
    Switzerland & $209.3$ & $3$:$18$ & $0$:$12$ & $16.5$\,x & $4\,315$ & $1.6$ \\
    London & $1\,368.1$ & $15$:$19$ & $0$:$42$ & $21.9$\,x & $20\,390$ & $6.0$ \\
    Madrid & $497.3$ & $1$:$22$ & $0$:$04$ & $17.0$\,x & $32\,293$ & $2.0$ \\
    \hline
  \end{tabular}
\end{table}

We evaluate query times in Table~\ref{tab:queries}.
We measured the average times for $10\,000$ queries with source and destination stop chosen uniformly at random.
For earliest arrival queries, the departure time was chosen uniformly at random on the first day; for profile queries, the departure time range is the entire first day.
We evaluated queries for three different variants:
The basic trip-based algorithm (TB), using prefix trees as described in section~\ref{sec:algorithm} (PT), and using both prefix and postfix trees as described in section~\ref{sec:splitting_trees} (ST).
The ST variant leads to larger query graphs than the PT variant.
This is to be expected, as some information gets lost in the transformation, and some prefixes may connect to more postfixes than required.
This does not affect correctness, because all optimal line sequences are still contained in the query graph.
It does, however, lead to increased query times for ST in comparison to PT.
Nevertheless, the time required to construct the query graph on the Germany instance is lower for ST, since the split trees contain fewer nodes in total than the original prefix tree.
Profile query times are much higher on the metropolitan networks than on the generally larger country-sized networks.
In part, this is because they are less structured than the larger networks, which leads to larger query graphs.
However, on the metropolitan networks, the set of optimal journeys is also much larger than on the others, which slows down the query algorithm.

\begin{table}[tp]
  \centering
  \caption{Query figures.
    Listed are the query graph size (nodes + edges),
    the time required to construct the query graph,
    and the time required to perform an earliest arrival and a 24h profile query.
    The first block refers to the basic trip-based algorithm, where no query graph is used.
    The second block uses a prefix tree for each source stop, as in section~\ref{sec:algorithm}.
    The third block uses the split trees for source and destination stop, as in section~\ref{sec:splitting_trees}.
  }\label{tab:queries}
  \begin{tabular}{l r r r r r}
    \hline
    \,       & \,    & Query graph & Query graph & EA & profile \\
    Instance & Var.\ & size [N+E] & time [{\textmu}s] & [{\textmu}s] & [{\textmu}s] \\\hline
    Germany & TB & --- & --- & $30\,856$ & $192\,952$ \\
    Sweden & TB & --- & --- & $2\,760$ & $16\,532$ \\
    Switzerland & TB & --- & --- & $1\,780$ & $18\,104$ \\
    London & TB & --- & --- & $1\,374$ & $96\,114$ \\
    Madrid & TB & --- & --- & $711$ & $54\,118$ \\
    \hline
    Germany & PT & $41+58$ & $994.4$ & $63.3$ & $155.0$ \\
    Sweden & PT & $23+32$ & $24.6$ & $40.4$ & $88.6$ \\
    Switzerland & PT & $38+59$ & $34.0$ & $45.8$ & $155.9$ \\
    London & PT & $91+196$ & $138.2$ & $101.1$ & $2\,786.6$ \\
    Madrid & PT & $150+407$ & $306.9$ & $81.7$ & $6\,913.8$ \\
    \hline
    Germany & ST & $124+232$ & $81.1$ & $75.0$ & $430.5$ \\
    Sweden & ST & $66+122$ & $32.5$ & $27.2$ & $207.1$ \\
    Switzerland & ST & $118+233$ & $76.1$ & $32.7$ & $327.6$ \\
    London & ST & $331+1242$ & $1\,583.3$ & $141.4$ & $14\,545.4$ \\
    Madrid & ST & $456+2073$ & $11\,822.9$ & $165.8$ & $28\,919.0$ \\
    \hline
  \end{tabular}
\end{table}

\begin{table}[tp]
  \centering
  \caption{Comparison with the state of the art.
    Results taken from \protect{\cite{Bast2014a,Bast2016,Bast2014,Delling2015,Strasser2014,Wang2015}}.
    Algorithms computing Pareto-optimal journeys with respect to the number of transfers in addition to arrival time are marked in column ``tr.''
    Profile queries are marked in column ``pr.''
  }\label{tab:comparison}
  \begin{tabular}{l l r r c c r r r}
    \hline
    algorithm & instance & stops & conn.& tr.& pr.& mem.\ & pre.\ & query \\
    \, & \, & [$10^3$] & [$10^6$] & & & [GB] & [h] & [{\textmu}s]\\\hline
    CSA~\cite{Strasser2014} & Germany & $252.4$ & $46.2$ & $\circ$ & $\circ$ & --- & --- & $298.6$\,k \\
    ACSA~\cite{Strasser2014} & Germany & $252.4$ & $46.2$ & $\circ$ & $\circ$ & n/a & $0.2$ & $8.7$\,k \\
    TP~\cite{Bast2014} & Germany & $248.4$ & $13.9$ & $\bullet$ & $\circ$ & $140.0$ & $372.0$ & $300.0$ \\
    Sc-TP~\cite{Bast2016} & Germany & $250.0$ & $15.0$ & $\bullet$ & $\circ$ & $1.2$ & $16.5$ & $32.0$\,k \\
    TB & Germany & $247.9$ & $27.1$ & $\bullet$ & $\circ$ & $23.2$ & $231.3$ & $156.1$ \\
    \rule{0ex}{1.5em}TTL~\cite{Wang2015} & Sweden & $51.4$ & n/a & $\circ$ & $\circ$ & $\approx0.5$ & $0.2$ & $\approx10.0$ \\
    PTL~\cite{Delling2015} & Sweden & $51.1$ & $12.7$ & $\bullet$ & $\circ$ & $12.3$ & $36.2$ & $27.6$ \\
    TB & Sweden & $50.7$ & $6.1$ & $\bullet$ & $\circ$ & $1.6$ & $3.8$ & $59.7$ \\
    \rule{0ex}{1.5em}PTL~\cite{Delling2015} & Switzerland & $27.1$ & $23.7$ & $\bullet$ & $\circ$ & $12.7$ & $61.6$ & $21.7$ \\
    TB & Switzerland & $27.8$ & $4.7$ & $\bullet$ & $\circ$ & $1.6$ & $2.7$ & $108.8$ \\
    \rule{0ex}{1.5em}CSA~\cite{Dibbelt2013} & London & $20.8$ & $4.9$ & $\circ$ & $\circ$ & --- & --- & $1.8$\,k \\
    PTL~\cite{Delling2015} & London & $20.8$ & $5.1$ & $\bullet$ & $\circ$ & $26.2$ & $49.3$ & $30.0$ \\
    TB & London & $20.8$ & $5.0$ & $\bullet$ & $\circ$ & $6.0$ & $11.6$ & $1.7$\,k \\
    \rule{0ex}{1.5em}TTL~\cite{Wang2015} & Madrid & $4.6$ & n/a & $\circ$ & $\circ$ & $\approx0.4$ & $0.1$ & $\approx30.0$ \\
    PTL~\cite{Delling2015} & Madrid & $4.7$ & $4.5$ & $\bullet$ & $\circ$ & $9.9$ & $10.9$ & $64.3$ \\
    TP~\cite{Bast2014a} & Madrid & $4.6$ & $4.8$ & $\bullet$ & $\circ$ & n/a & $185.0$ & $3.1$\,k \\
    TB & Madrid & $4.6$ & $5.3$ & $\bullet$ & $\circ$ & $2.0$ & $1.1$ & $12.0$\,k \\
    \hline
    \rule{0ex}{1.5em}ACSA~\cite{Strasser2014} & Germany & $252.4$ & $46.2$ & $\circ$ & $\bullet$ & n/a & $0.2$ & $171.0$\,k \\
    TP~\cite{Bast2014} & Germany & $248.4$ & $13.9$ & $\bullet$ & $\bullet$ & $140.0$ & $372.0$ & $5.0$\,k \\
    TB & Germany & $247.9$ & $27.1$ & $\bullet$ & $\bullet$ & $23.2$ & $231.3$ & $511.6$ \\
    \rule{0ex}{1.5em}PTL~\cite{Delling2015} & Sweden & $51.1$ & $12.7$ & $\circ$ & $\bullet$ & $0.7$ & $0.5$ & $12.1$ \\
    TB & Sweden & $50.7$ & $6.1$ & $\bullet$ & $\bullet$ & $1.6$ & $3.8$ & $239.6$ \\
    \rule{0ex}{1.5em}PTL~\cite{Delling2015} & Switzerland & $27.1$ & $23.7$ & $\circ$ & $\bullet$ & $0.7$ & $0.7$ & $24.5$ \\
    TB & Switzerland & $27.8$ & $4.7$ & $\bullet$ & $\bullet$ & $1.6$ & $2.7$ & $403.7$ \\
    \rule{0ex}{1.5em}PTL~\cite{Delling2015} & London & $20.8$ & $5.1$ & $\circ$ & $\bullet$ & $1.3$ & $0.9$ & $74.3$ \\
    CSA~\cite{Dibbelt2013} & London & $20.8$ & $4.9$ & $\bullet$ & $\bullet$ & --- & --- & $466.0$\,k \\
    TB & London & $20.8$ & $5.0$ & $\bullet$ & $\bullet$ & $6.0$ & $11.6$ & $16.1$\,k \\
    \rule{0ex}{1.5em}PTL~\cite{Delling2015} & Madrid & $4.7$ & $4.5$ & $\circ$ & $\bullet$ & $0.4$ & $0.4$ & $111.9$ \\
    TB & Madrid & $4.6$ & $5.3$ & $\bullet$ & $\bullet$ & $2.0$ & $1.1$ & $40.7$\,k \\
    \hline
  \end{tabular}
\end{table}

We compare variant ST, using prefix and postfix trees, to other state of the art algorithms in Table~\ref{tab:comparison}.
Algorithms based on labelling approaches are generally the fastest.
In particular, for single criterion queries, they dominate other preprocessing-based approaches with regard to query times, preprocessing time and memory consumption.
PTL~\cite{Delling2015} supports multi-criteria queries, at the cost of massive increases in both preprocessing time and memory consumption, while TTL~\cite{Wang2015} only performs single-criterion queries.
TP~\cite{Bast2010,Bast2016,Bast2014} can answer bicriteria profile queries in a few milliseconds, even on large networks.
The original TP had the drawbacks of very long preprocessing times and a large memory consumption.
More recently, Scalable Transfer Pattern~\cite{Bast2016} has made impressive improvements on this front, at the cost of increased query times.

On the metropolitan networks, our algorithm performs notably worse than could be expected, although query times are still in the low milliseconds.
As previously mentioned, this is mostly due to the much higher number of journeys compared to the country-sized networks.
For bicriteria queries on the country-sized networks, our algorithm has preprocessing costs one order of magnitude less than PTL, while query times are similar.
Note, however, that PTL has not been evaluated for bicriteria profile queries, making direct comparison difficult.
In comparison to Scalable TP, our query times are two orders of magnitude lower, at the cost of one order of magnitude for preprocessing costs.
As such, our algorithm enables the currently fastest bicriteria profile queries on large realistic instances, with reasonable preprocessing overhead.
On very large instances, such as Germany, preprocessing time and memory consumption may be prohibitive for some use cases.
This is a subject of future research.

\section{Conclusion}\label{sec:conclusions}
We introduced a speed-up technique for the basic trip-based public transit routing algorithm~\cite{Witt2015}.
This technique applies principles sharing some similarities to those behind Transfer Patterns~\cite{Bast2010,Bast2014} and Hub Labelling~\cite{Cohen2003} to the trip-based model and expands on them.
The resulting algorithm enables query times on the microsecond scale on large realistic public transit networks with moderate preprocessing cost, occupying a Pareto-optimal spot among current state of the art algorithms.

Future work includes the study of different methods for cut node selection, with the goal of further reducing memory consumption and query graph size, developing tailored query algorithms to speed up queries on metropolitan networks, and making preprocessing more scalable by avoiding the computation of full one-to-all queries for all stops.
We are also interested in adapting this speed-up technique to different scenarios, such as other and/or more criteria, and stop-based routing.

\appendix
\bibliographystyle{hplain}
\bibliography{references}

\end{document}